\documentclass[twocolumn,floats,aps]{revtex4}
\usepackage{times,amsmath,psfrag,latexsym,pstricks,graphics}
 
\begin{document}
\title{Segregation and Stability of Binary Granular Mixtures}
\author{A.G. Swartz, J.B. Kalmbach, J. Olson and R.J. Zieve}
\affiliation{Physics Department, University of California at Davis}
\begin{abstract}
We measure stability of two-dimensional granular mixtures in a
rotating drum and relate grain configurations to stability.  For our
system, the smaller but smoother grains cluster near the center of the
drum, while the larger, rougher grains remain near the outer edge.
One consequence of the size segregation is that the smaller grains
heavily influence the stability of the heap.  We find that the maximum
angle of stability is a non-linear function of composition, changing
particularly rapidly when small grains are first added to a homogeneous
pile of large grains.  We conclude that the grain configuration within
the central portion of the heap plays a prominent role in stability.
\end{abstract}
\maketitle

Granular materials have received much attention due to their important
practical applications in fields ranging from geology to the food industry
\cite{Duran}.  Despite the simple behavior of individual grains, a
granular system can display complex collective behavior.  In different regimes
of density and motion, granular matter may most mimic a solid, liquid, or gas
\cite{Jaeger}.  Although a good deal of effort has gone into cataloging these
phases, there is little understanding of how the microscopic grain
configuration leads to specific macroscopic behavior; that is, which features
in an arrangement of grains are most important. Observed hysteresis in the
transition to crystalline order \cite{Daniels} and in the numbers of contacts
between grains \cite{Deboeuf} demonstrates that the microscopic configuration
contains subtle information on the system's history.  A more striking
illustration of the importance of microstructure comes from sound propagation
experiments, where a fractional size change in a single grain can change the
total sound transmission through a granular heap by up to 25\% \cite{Liu}.

Understanding the connection between configuration and behavior is particularly
necessary for moving from experimental studies of homogeneous mixtures to the
heterogeneous mixtures that appear in most industrial and other applications. 
Differences among grain configurations can be more significant in a mixture
than in a homogeneous sample.  Because the configurations vary more widely, the
microscopic arrangements may also have more effect on the overall behavior.
Conveniently, the increased variation in heterogeneous systems can also make
it easier to address the connection between the microscopic and macroscopic
regimes.

Heterogeneous mixtures commonly undergo segregation, a phenomenon of great
interest for materials processing \cite{Ottino}.  Segregation of binary
mixtures also occurs in the laboratory, in both vibration \cite{Krishna, Huerta}
and rotating drum experiments \cite{Lai, Puri, Prigozhin, Taberlet, Newey,
Yanagita, Peratt}, and can be caused by differences in particle size, shape, smoothness, or
density.  If a particular region of a granular heap plays the major role in
some macroscopic behavior, then a mixed heap should behave most like the
component that is heavily represented in the region.  Here we discuss
measurements on binary mixtures in a two-dimensional rotating drum.  We
describe our apparatus and the segregation patterns we observe.  We then
combine the segregation observations with measurements of avalanches  
to probe how different configurations influence stability.

\begin{figure}[b]
\includegraphics{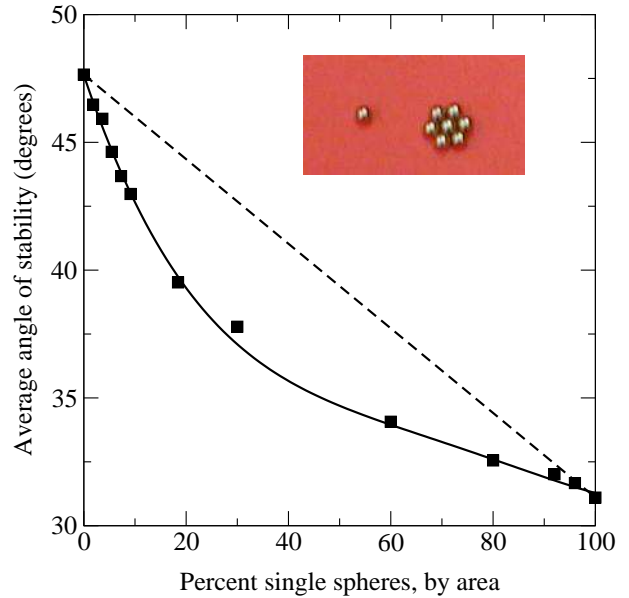}
\caption{Maximum stability angles for different pile compositions.  The solid
line is a guide the eye, and dotted line is the interpolation between the
values for homogeneous piles.  Inset: sphere and hexagon.}
\label{f:angles}
\end{figure}

The grains of our mixture, shown in the inset to Figure \ref{f:angles},
are composed of $\frac18$"-diameter ball
bearings.  We use both single spherical bearings and hexagonal shapes
created by welding 7 single steel ball bearings together, as described
elsewhere \cite{Ivan}.  The hexagonal shapes can form near-perfect
triangular lattices of the individual spheres, in which the joints
between balls are impossible to distinguish, demonstrating that the
welding process does not significantly distort the balls.

We use a rotating drum consisting of a $\frac18$"-thick aluminum sheet with a
13.96"-diameter circle cut out of it.  Two $\frac12$"-thick sheets of Plexiglas
sandwich the aluminum sheet and contain the mixture.  Confining the grains to a
single layer allows us to track visually the exact configuration of the grains
during the rotation. In general, only the top few layers of grains flow when a
critical angle of stability is reached while the bulk of the mixture remains
stable.  Extensive three-dimensional segregation experiments and models
\cite{Taberlet, Newey, Yanagita} have reported that binary mixtures will
segregate axially into bands.   Working in two dimensions eliminates
any possibility for axial instability.

The drum rotates about its center on an axle with rotational speed controlled
by a motor.  We rotate at 500 $\mu$Hz, or one full turn in just over half an
hour.  This slow speed allows for static limit behavior, characterized
by discrete avalanches. The rotation of the drum during an avalanche is about
one-third of a degree, for most purposes small enough to be ignored. 
We also verified experimentally that, within the uncertainty of our
measurements, the distribution of avalanches in angle does not change
with rotation speed up to 1 mHz.

The binary mixture is inserted into the tumbler through an opening in one side
of the aluminum sheet.  The opening is closed during rotation.  Since we
are studying the behavior of binary mixtures, the hexagonal shapes must
be entered carefully to avoid breakage.  As described below, even a small
percentage of single spheres in a pile of hexagonal grains noticeably
decreases the pile's stability, and broken hexagons have a similar effect.
After completing measurements on a pile, we check for broken shapes.
If more than 1\% of the hexagonal shapes are broken, we discard the data
and remeasure that composition.

Another method is to disassemble the tumbler and remove one of the
Plexiglas faces.  The shapes can then be placed in the aluminum sheet with
little risk of breaking.  However, if the screws which hold the Plexiglas
and the aluminum sheets together are not tightened equally, we find a
periodicity in avalanche behavior that reflects the rotation rate of the
drum. This arises from slight differences in how uniformly the grains are
confined to a single layer.  To avoid such bias, we adjust the container
until the rotation periodicity is not reflected in our data, and then
leave the container undisturbed through the entire set of measurements.

For each run the pile fills about 28\% of the container. Once a mixture is
loaded into the tumbler, we rotate at 500 $\mu$Hz for at least 20 minutes
before beginning to record avalanches.  This avoids anomalous results arising
from the configuration of the grains upon loading. Instead, we consider only
avalanches in a ``steady state," where each grain configuration has developed
from previous avalanches. We then rotate the drum for one to two hours to
obtain about 100 discrete avalanches, filming the entire time with
a digital video camera.  Avalanches are larger and less frequent for
the hexagons than for the spheres, so the filming takes correspondingly
longer. The frames taken immediately before and after each avalanche are
transfered to a computer.  The angle of stability before an avalanche
and the angle of repose afterward are calculated from the average slope of
the pile's free surface.  The pile's packing fraction before and after
each avalanche is also computed, as described elsewhere \cite{Jeremy}.

\begin{figure}[t]
\pspicture(0,0)(8,6)
\rput[lb](0,2.75){\scalebox{.21}{\includegraphics{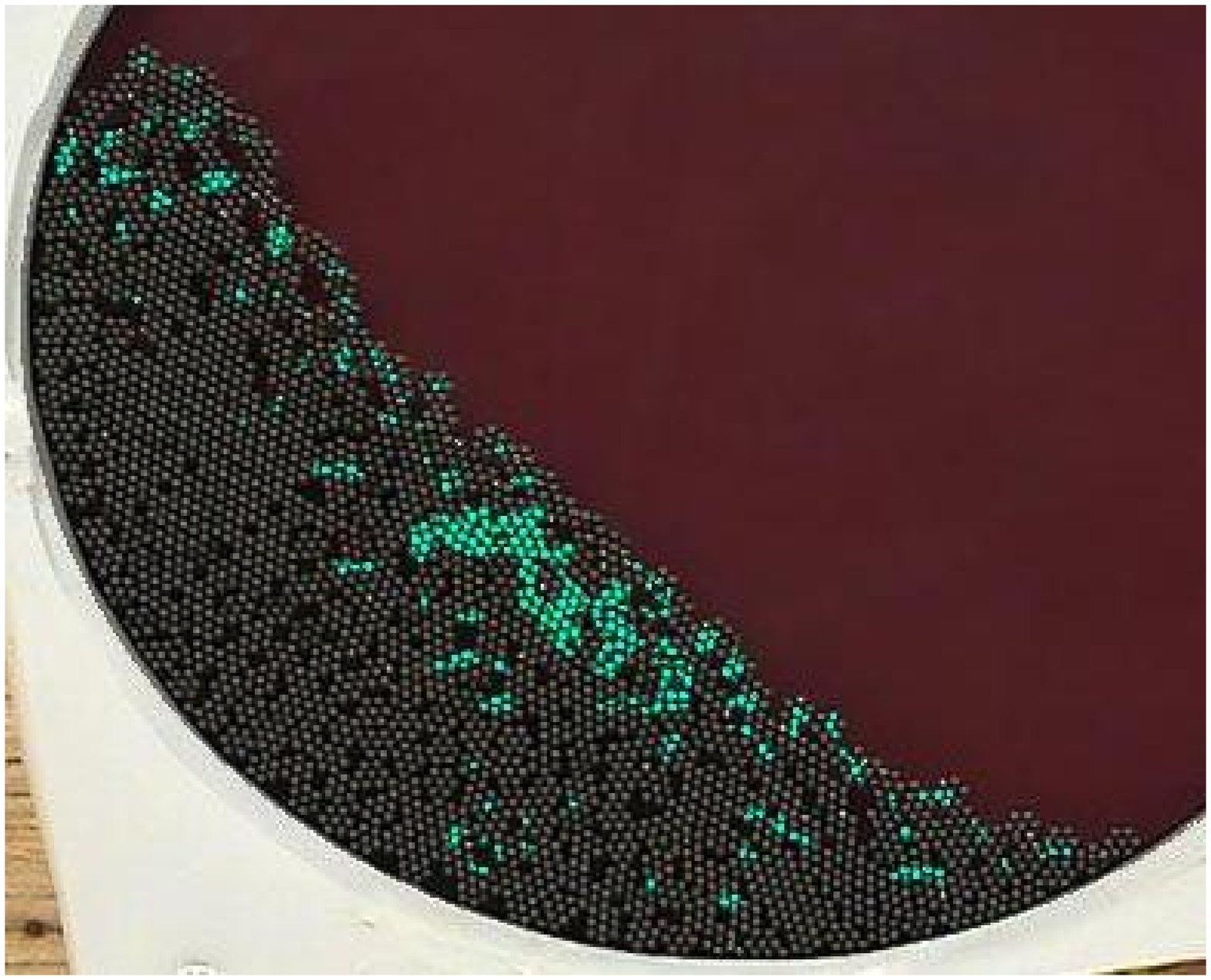}}}
\rput[lb](4.2,2.75){\scalebox{.21}{\includegraphics{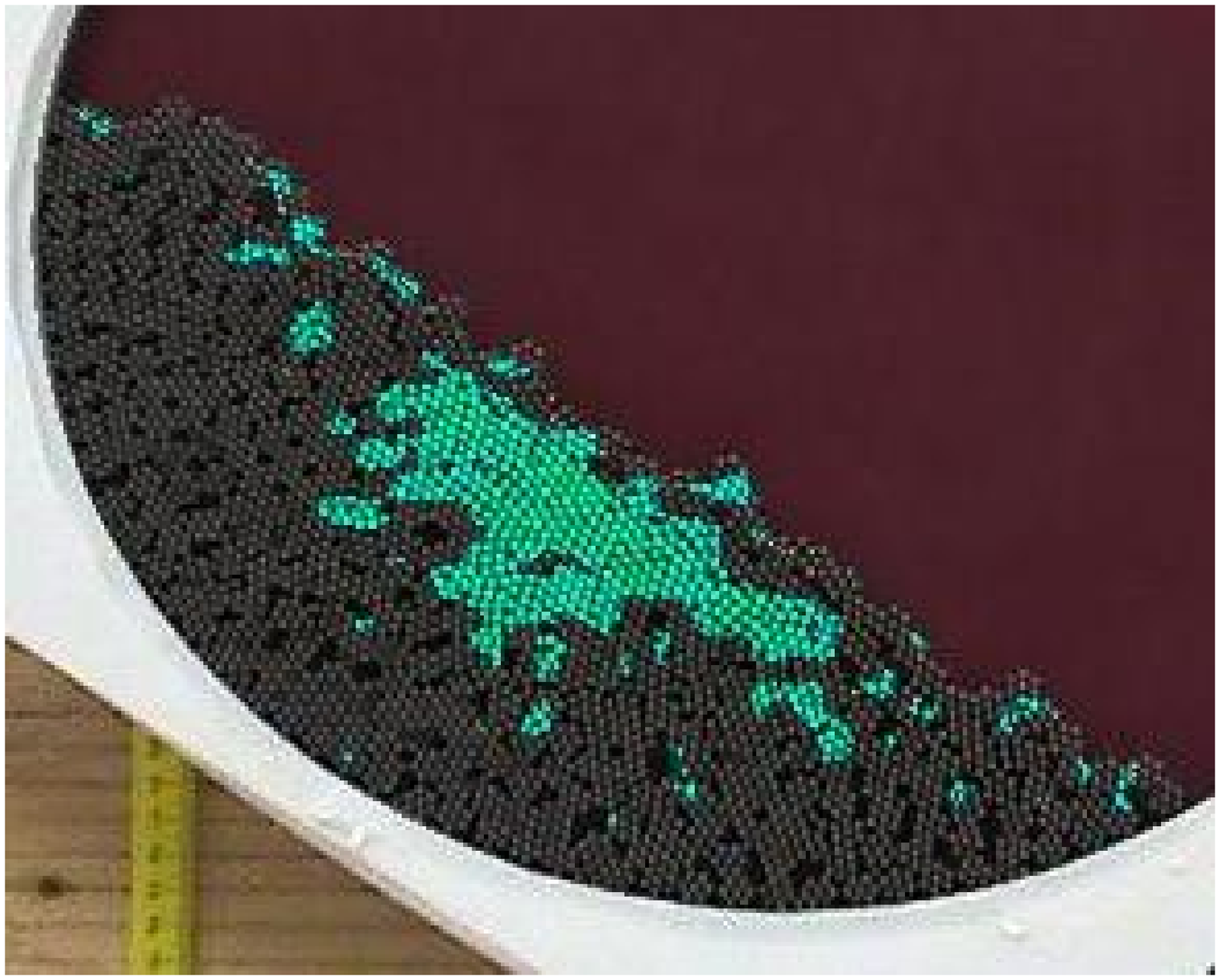}}}
\rput[lb](0,0){\scalebox{.21}{\includegraphics{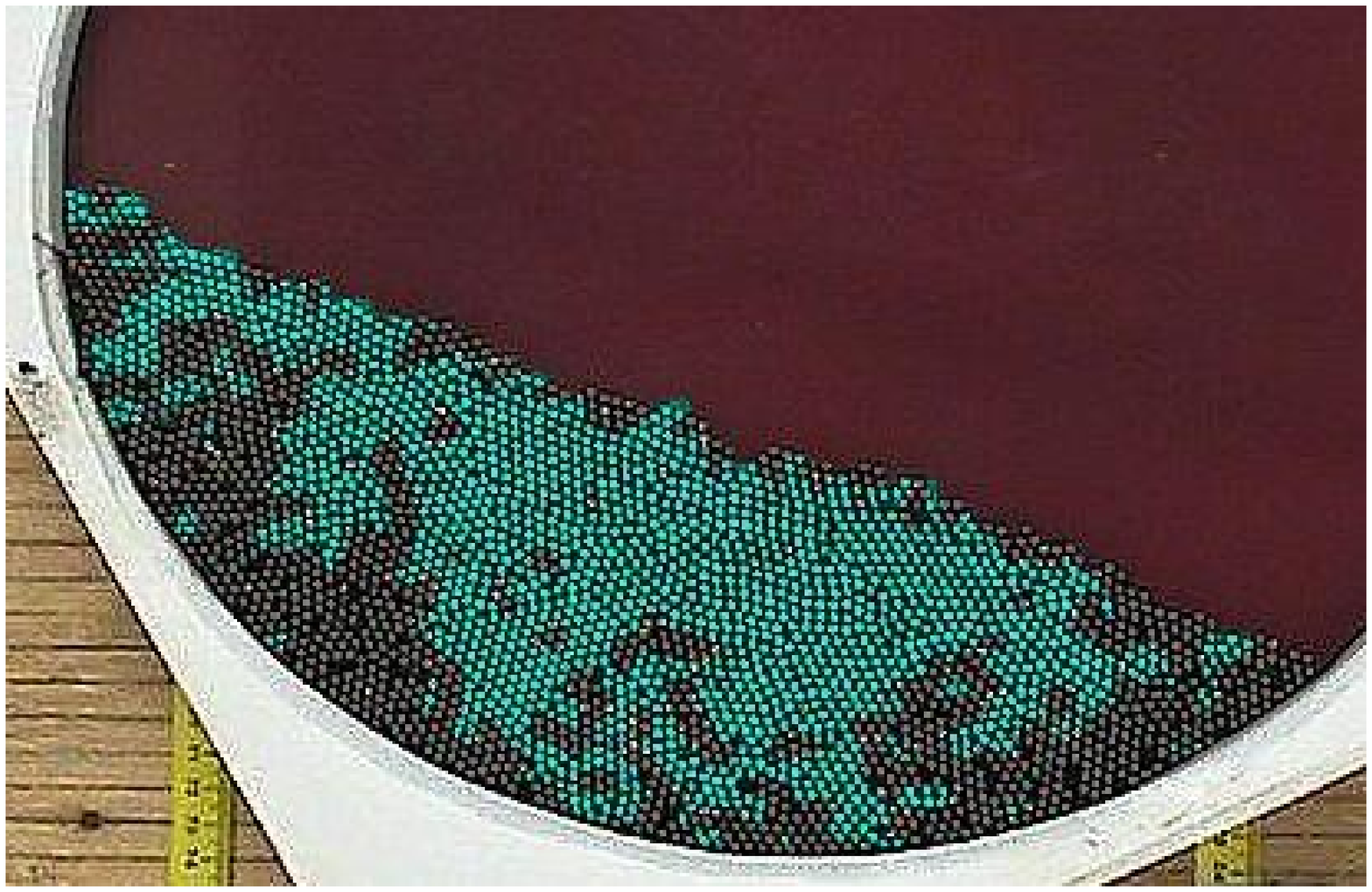}}}
\rput[lb](4.2,0){\scalebox{.21}{\includegraphics{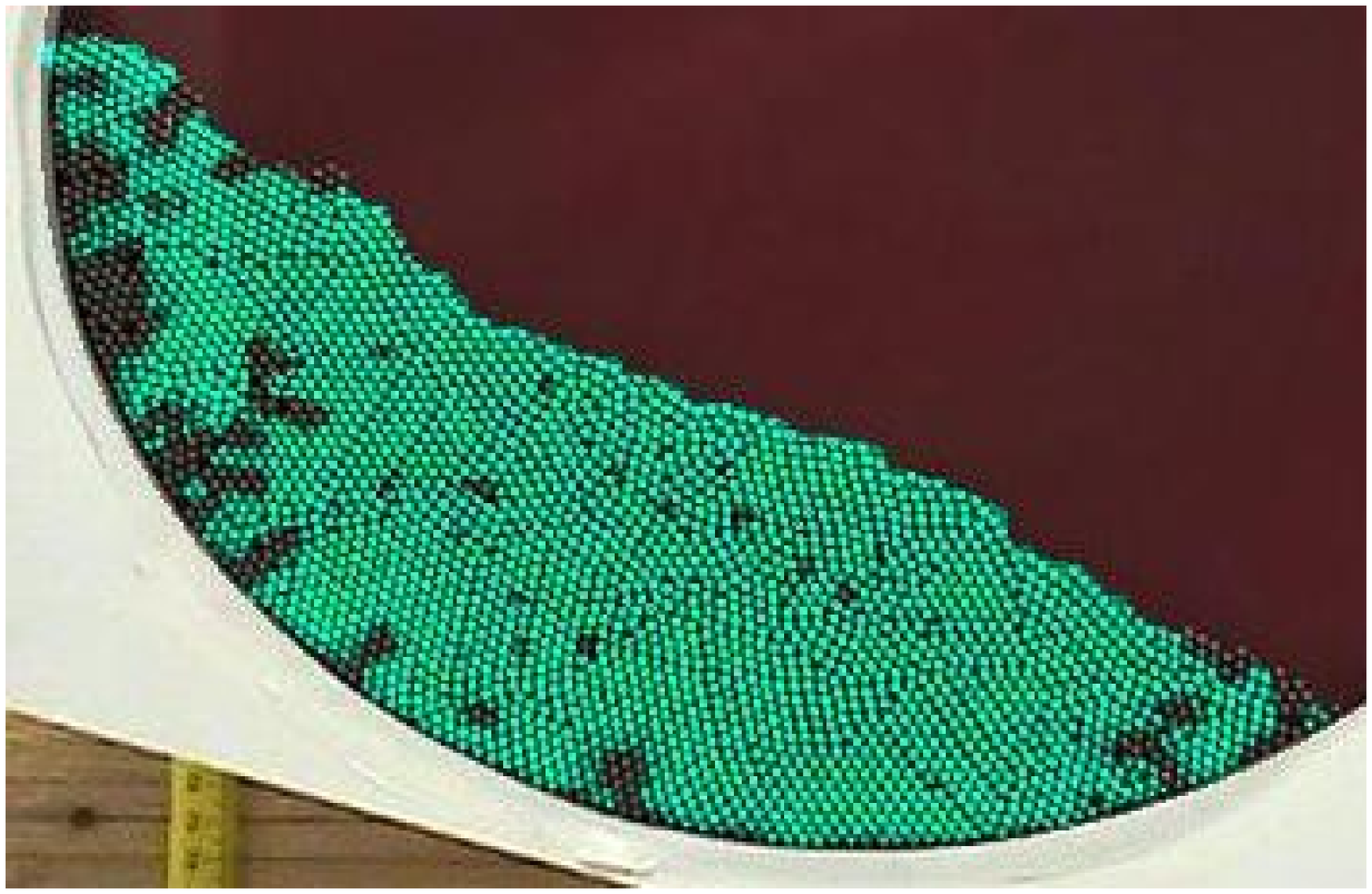}}}
\rput[lb](3.5,5.5){\scalebox{1.1}{\white a}}
\rput[lb](7.7,5.5){\scalebox{1.1}{\white b}}
\rput[lb](3.5,2.1){\scalebox{1.1}{\white c}}
\rput[lb](7.7,2.1){\scalebox{1.1}{\white d}}
\endpspicture
\caption{Progression of segregation for mixture of hexagons (gray)
and spheres (green).  Shown are mixtures with
(a) 7\%, (b) 18\%, (c) 58\%, and (d) 89\% single spheres by area.}
\label{f:segregate}
\end{figure}

For many of the measurements we use brass ball bearings for the single balls,
which we can distinguish by color from the steel hexagons.  Since the brass
balls weigh about 10\% more than the steel balls, the change in material could
affect the segregation and avalanche behavior. For a few mixtures we also
performed measurements with all steel balls, and verified that substituting the
brass ball bearings does not change the avalanche angles and packing
fractions.  In Figure \ref{f:segregate}, we show typical configurations of the
mixtures.  In the original images, which are included in an appendix, about 70\% of the brass balls
show unambiguously from their yellowish color.  Since light reflected from
nearby brass balls or red background can give even the steel balls a slight
yellow tint, distinguishing the remaining single brass balls is more
difficult.  To better convey the configurations in the small images of
Figure \ref{f:segregate}, the brass balls are colored green for
visibility.  We select those balls with the strongest yellow tint as brass,
with the cutoff chosen so that the correct number of balls are identified as
brass.  A few steel balls with brass
neighbors fall in the ``brass" group and some isolated brass balls do not.
Nonetheless, the bulk of the brass balls are selected accurately and the
general patterns shown are correct.

The spheres cluster near the center of the drum, usually
slightly below the free surface of the heap. The hexagons tend to
inhabit the outer portions of the drum.  Several previous studies
\cite{Prigozhin,Taberlet,Newey}, both numerical and experimental, have
also found that size differences lead to radial segregation, with smaller
shapes moving to the drum center.  Interestingly, other work \cite{Lai,
Puri} shows that friction can also cause radial segregation, with rougher
shapes moving to the center.  The size and roughness characteristics can
compete with each other.  When a mix of small smooth grains and large
rough grains are dropped onto a pile, the grain types form striations
within the pile \cite{Makse}.  A much simpler segregation occurs when
the smaller grains are also rougher.  In this case the larger shape
collects at the bottom and the smaller shape at the top of the pile,
with a single sharp demarcation between the two regions \cite{Makse}.
Geologists have long been aware of such patterns, using striations
to understand how sediment settled out of the flows that created rock
structures \cite{Gray}. For our shapes, the hexagons are clearly rougher,
but their large size apparently overwhelms their roughness in the
segregation process.  We find no striations or fingering, but rather a
clustering consistent with pure size segregation.  Possibly this behavior
would change for large rotation speeds, where the free surface is steepest
in the center and the shape with greater repose angle collects there.
At our slow speeds, with a near-linear free surface, the reduced influence
of roughness is not surprising.

Figure \ref{f:segregate} shows the progression of the segregation as
the concentration of singles increases.  For small percentages (2-8\%)
the singles, which converge towards the bottom of the laminar layer,
group themselves into small clusters.  The size and number of these
clusters increases with composition.  The 7\% composition (Figure
\ref{f:segregate}a)
shows singles congregating into these distinct clumps of 5 to 10 singles.
These clusters all lie within an elongated area running roughly parallel
to the free surface.

At 18\% spheres (Figure \ref{f:segregate}b), the small clusters have grown together and the
singles have saturated the elongated area near the center of the free surface. 
With further increase in the fraction of single spheres, this central region
grows, spreading outward both along the free edge and perpendicular to it.  At
the other extreme, in mixtures with primarily single spheres, the few remaining
hexagons lodge along the outer edge of the container.

We now turn to the steady-state averages of the maximum  angle of stability for
our binary mixtures.  From previous work \cite{Jeremy} we know that hexagons
are significantly more stable than single ball bearings.  Similar results were
reported in \cite{Cantelaube} for disks and pentagons.  As shown in Figure
\ref{f:angles}, varying the fraction of spheres in the mixture changes the
stability of the pile in a decidely nonlinear manner.  For mixtures with a
small percentage of spheres, the stability angle decreases nearly three times
as quickly as  the linear interpolation between the angles for homogeneous
piles. At the other extreme, a few hexagons in a pile of spheres have almost no
effect on the pile's stability.

\begin{table}[bt]
\caption{Locations of single spheres for avalanches with angles at least
one standard deviation from the mean.}
\label{t:distribution}
\begin{tabular}{l||l|l}
& Highest angles & Lowest angles\\
\hline
Through & 0 & 6\\
Near & 4 & 5\\
Far & 10 & 2\\
\hline
Total & 14 & 13
\end{tabular}
\end{table}

In light of the segregation behavior described above, one explanation
of the observed angles of stability
is that most avalanches are triggered from the central region of the drum,
which has a disproportionately large concentration of single spheres.
Conversely, a small number of hexagons in a pile of spheres move to the
outside of the drum and have almost no effect on the avalanche behavior.

\begin{figure}[h]
\scalebox{.35}{\includegraphics{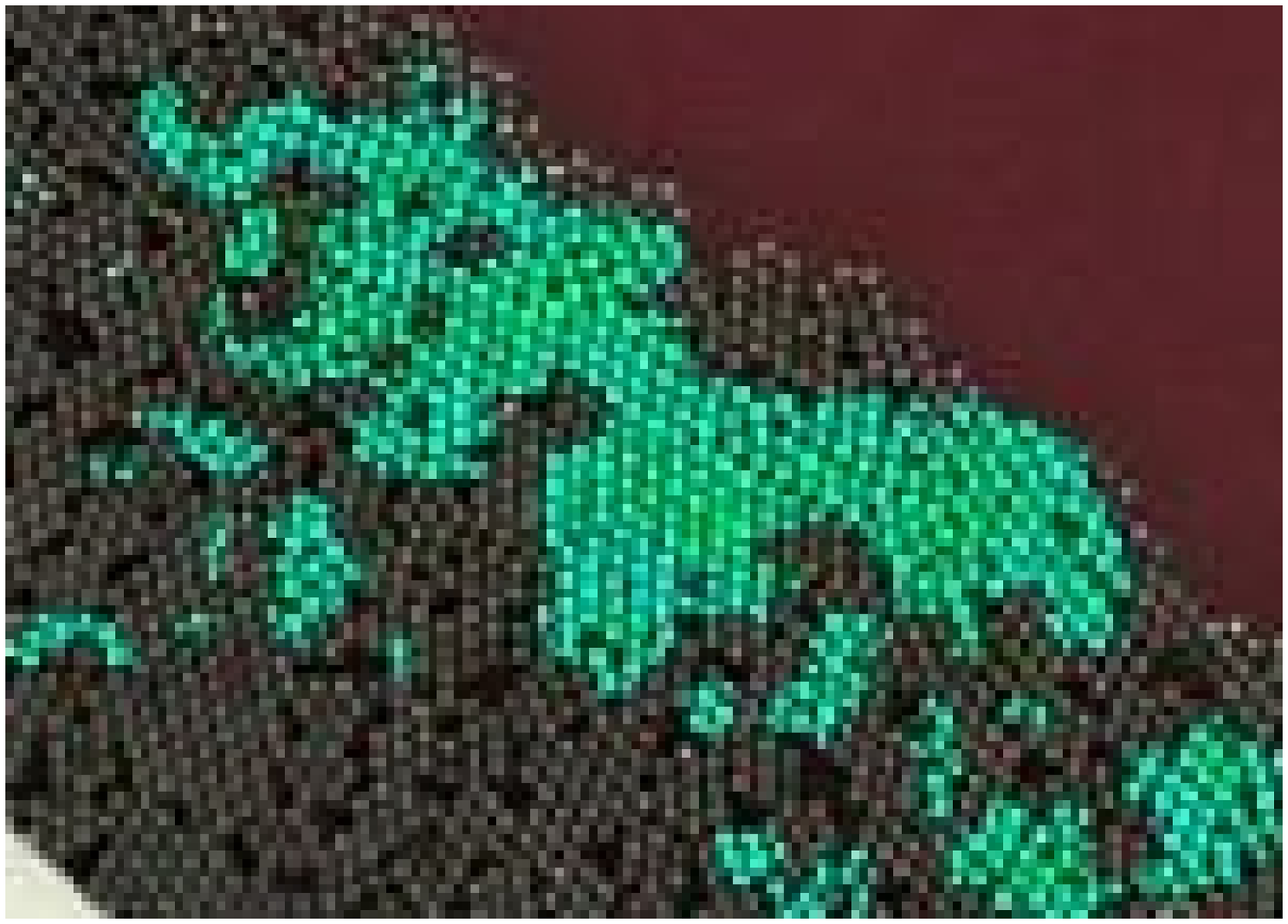}}\vspace{.03in}\\
\scalebox{.35}{\includegraphics{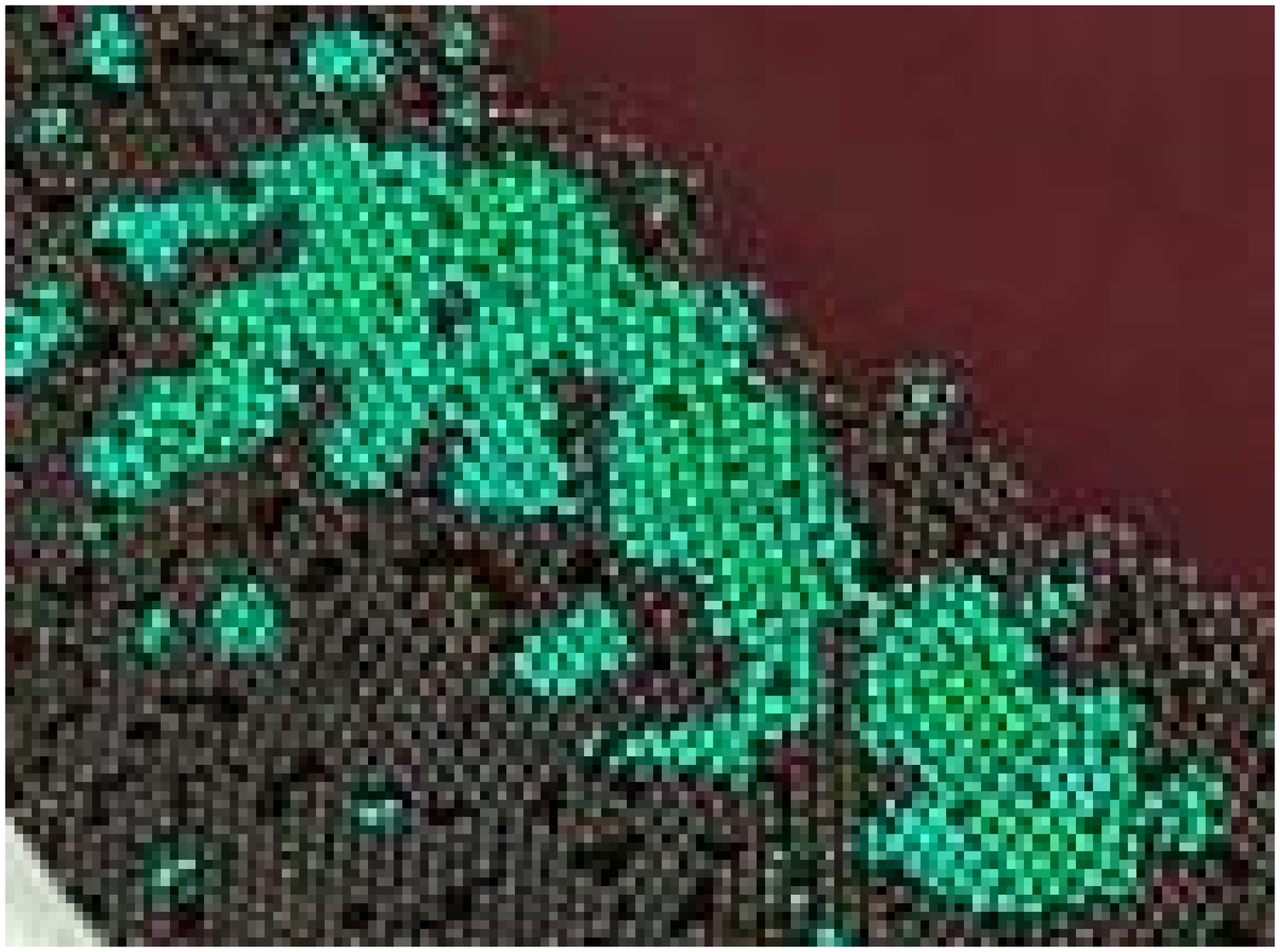}}\vspace{.03in}\\
\scalebox{.35}{\includegraphics{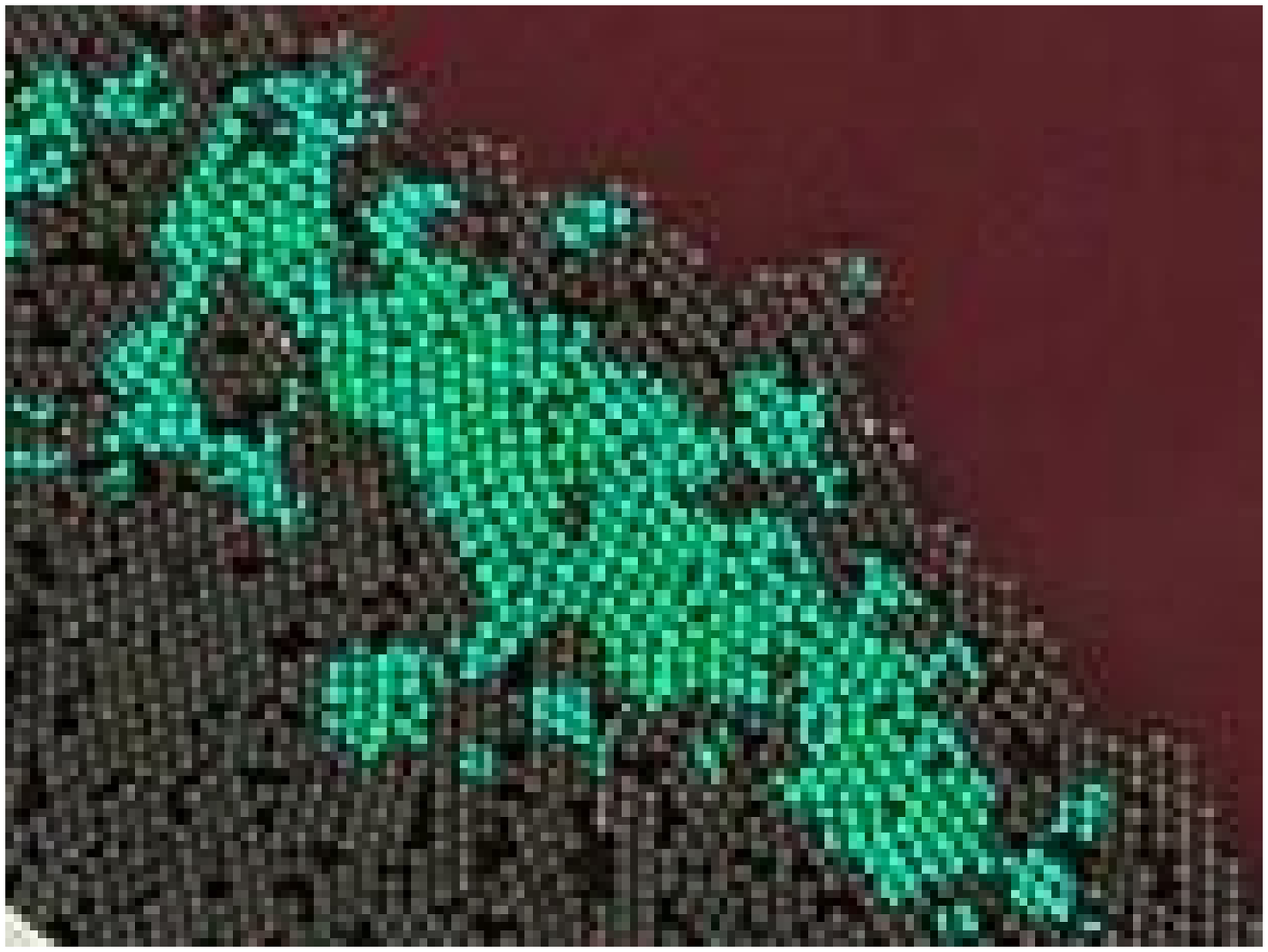}}
\caption{Examples of the arrangements of single spheres near the free
surface of the pile.  We classify these as ``through" (top), ``close"
(center), and ``far" (bottom) from the surface.}
\label{f:nearfar}
\end{figure}

The individual arrangements just before each avalanche make the influence of
the central region even more apparent.  Figure \ref{f:nearfar} shows three
configurations for the pile with   18\% spheres.  In the upper image the
central clump of single  spheres reaches through to the free surface of the
pile over a length of several grains.  In the middle image the spheres track
the free surface, with only a single row of hexagons atop the spheres. In the
lower image the single spheres are farther from the boundary, and the few that
approach it most closely are not part of the main clump of spheres. Although
categorizing a configuration as ``through," ``close," or ``far" is subjective,
two viewers agreed on over 90\% of the
images.

For the 18\% pile, we examined the images for all avalanches that occurred at
angles at least one standard deviation from the mean, with the results in Table
\ref{t:distribution}.  A clear distinction emerges despite the subjective
nature of the analysis.  In all but two of the 13 lowest-angle avalanches, the
clump 
of single spheres extended through to the free surface or came 
very close.  By contrast, for the 14 highest-angle avalanches, the singles
never reached the free surface and came close in only four cases. Thus single
balls within the top few layers have the greatest impact in triggering
avalanches.  We find similar trends for other single-hexagon mixtures, although
the 18\% is the most illuminating. For lower fractions of single spheres the
central clump disperses into small clusters and ``close" and ``through"
configurations become very rare.  For  higher concentrations, the central clump
of singles is larger and nearly always reaches the free surface. In both cases
the contrast between the high-angle and low-angle avalanches recedes.

\begin{figure}[t]
\scalebox{.45}{\includegraphics{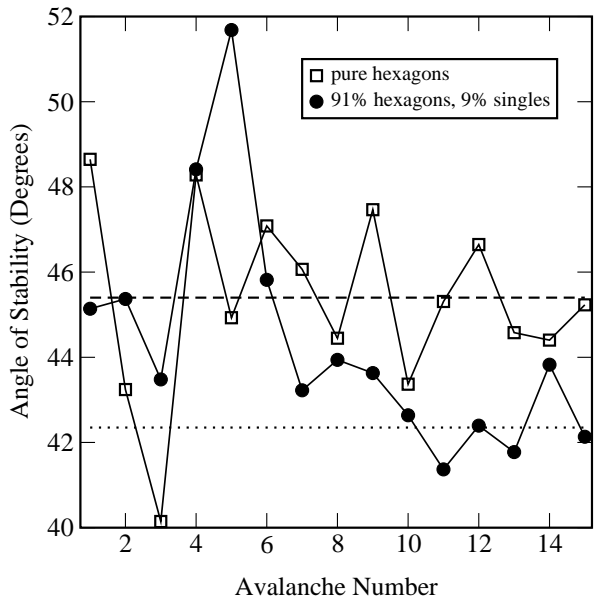}}
\caption{Avalanche angle for the first several avalanches for homogeneous
hexagons (open squares) and a mixture with 9\% singles (filled circles).
Dashed and dotted lines are averages of the final six avalanches for
the pure hexagon and the mixture, respectively.}
\label{f:initial}
\end{figure}

As a further test, we examined stability of configurations far from the typical
segregation patterns.  Using a mixture of hexagons and spheres, we first loaded
the spheres into the drum, so that they clustered along the bottom edge of the
drum, and placed the hexagons on top of them.  This starts the spheres as far
as possible from their eventual position in the center just below the free
surface.  Therefore, for the first few avalanches the singles are peripheral
and we expect to see stability comparable to the initial stability of a
homogeneous hexagonal pile.  During these first few avalanches, the singles
rotate with the drum until they reach the top of the pile and begin
participating in the avalanches.  Next we expect a transition period in which
the less stable spheres are no longer extraneous and the stability angle should
decrease.  Once the singles have migrated to their usual central position,
subsequent avalanches should reflect the usual characteristics of the steady
state.

Figure \ref{f:initial} shows this test of initial stability for a mixture
with 9\% spheres, as well as a similar measurement of initial avalanches for a
homogeneous pile of hexagons.  Because of the scatter in individual avalanches,
we measure the stability angles for the first 15 avalanches in several trials
at each composition.  Between trials the drum is emptied and reloaded.  The
data of Figure \ref{f:initial} are averaged across these trials, avalanche by
avalanche.  Even for a homogeneous pile, the first few avalanches depend on the
packing produced by loading the drum rather than on configurations resulting
from previous avalanches, and show wide variations in the maximum angle before
the avalanche.  The hexagonal pile serves  as a control, testing how
configurational differences other than  segregation affect the stability
angles.  The first six avalanches of both compositions have average angle of
stability similar to that of the homogeneous hexagons.  This indicates that,
initially, the singles are indeed peripheral both physically and in their
effect on the avalanches. After about 10 avalanches, the average stability for
the 9\% composition has transitioned towards the steady-state average. 
Visually, this corresponds well to when the single shapes permeate the central
area of the mix and start the clumping characteristic of radial segregation. 
The segregation occurs within about $\frac13$ of a full rotation.

In conclusion, we have measured stability for various mixtures of a small
smooth shape and a large rough shape.  We find that radial segregation  in our
system is governed primarily by size difference rather than roughness, with the
smaller grains moving to the center of the container.  Measuring stability, we
find nonlinear behavior characterized by a sharp decrease in stability when a
small amount of a low-stability shape is introduced to a homogeneous system. 
We further show that the nonlinear behavior in stability angle as a function of
composition depends on the radial segregation pattern.

We plan to extend this work in several ways.  By using faster rotation speeds
we can observe dynamical angles of stability.  It will be particularly
interesting to see whether the rougher hexagonal shapes still segregate to
the outside of the container at high speeds.  We will also examine in
more detail how the segregation occurs, and look at other grain shapes.

This work was supported in part by the National Science Foundation's Research
Experience for Undergraduates Program under PHY-0243904.

\begin{center}
APPENDIX
\end{center}

\begin{figure}[t]
\scalebox{.4}{\includegraphics{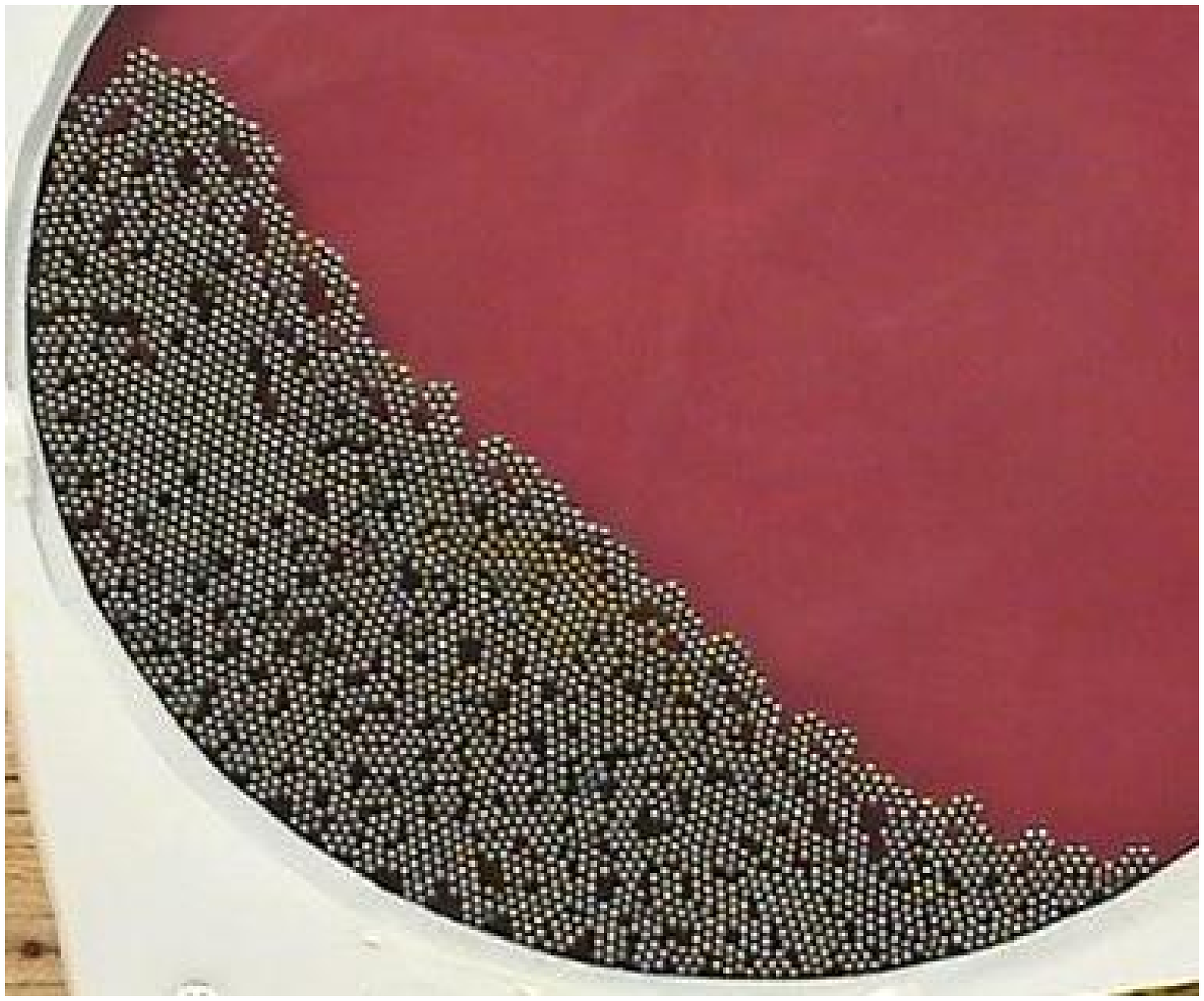}}\vspace{.06in}\\
\scalebox{.4}{\includegraphics{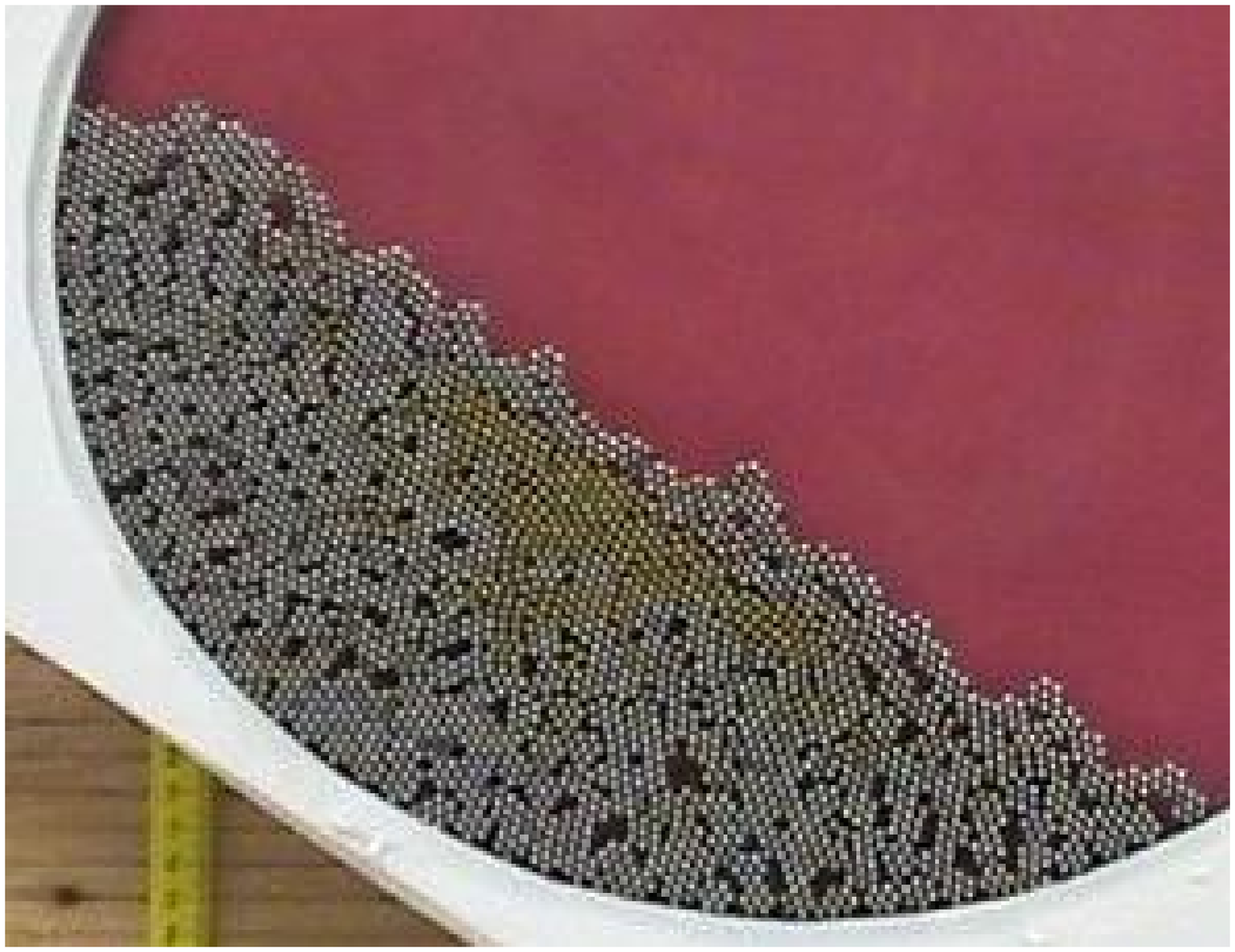}}\vspace{.06in}\\
\scalebox{.4}{\includegraphics{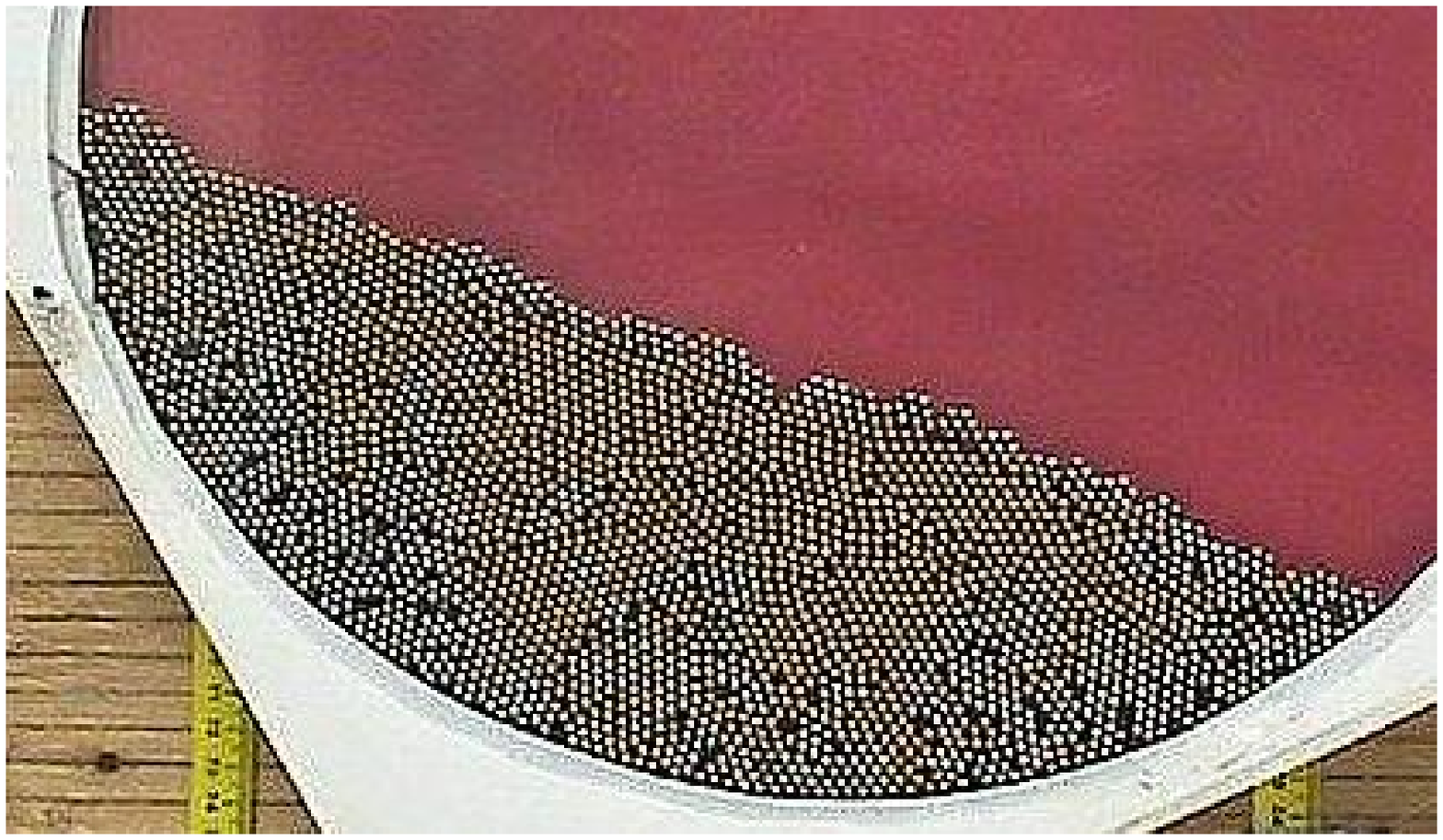}}\vspace{.06in}\\
\hspace*{.2in}\scalebox{.4}{\includegraphics{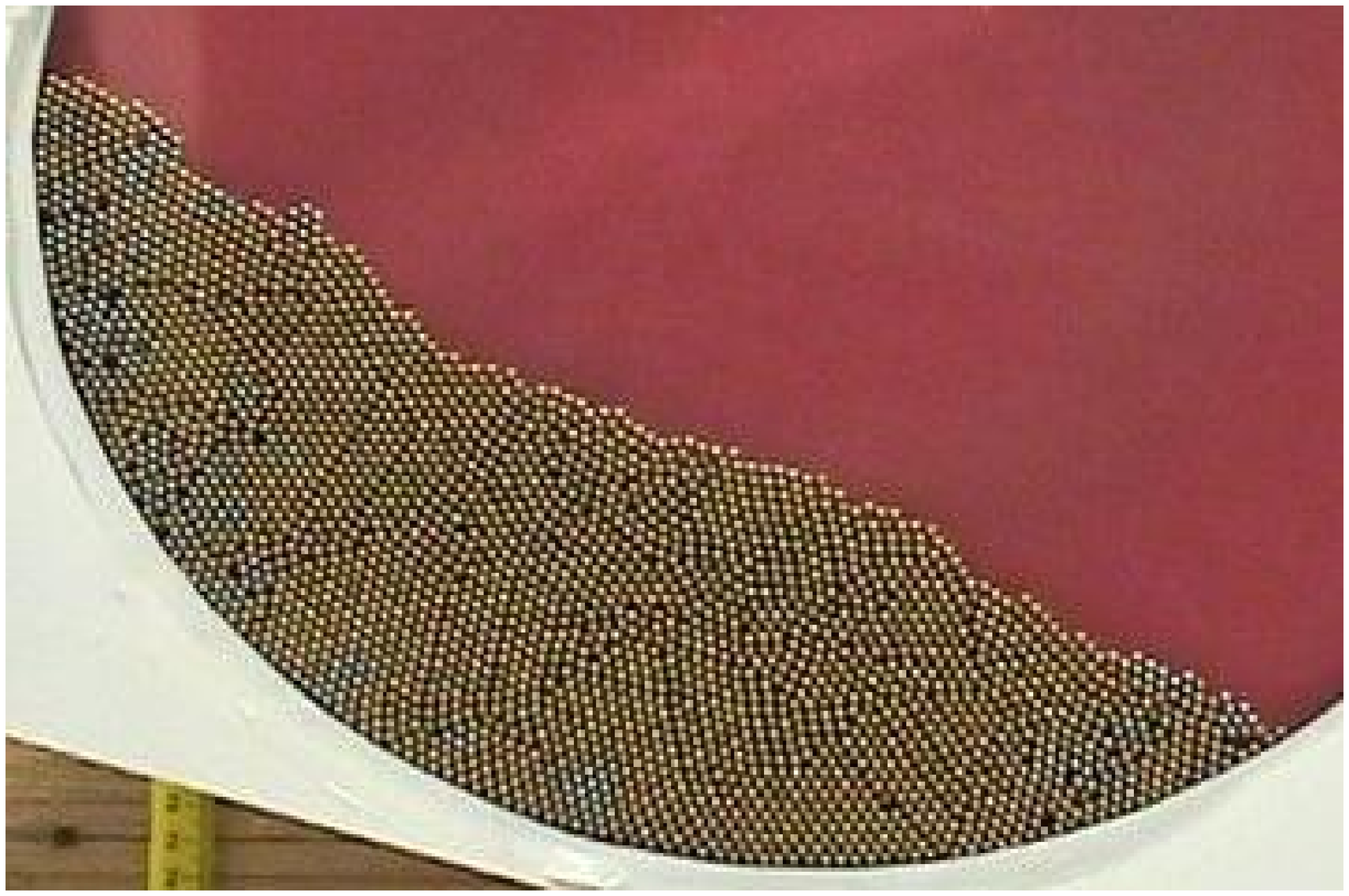}}
\caption{Images of Figure \ref{f:segregate} with original colors.}
\label{f:segcolor}
\end{figure}

\begin{figure}[tbh]
\scalebox{.35}{\includegraphics{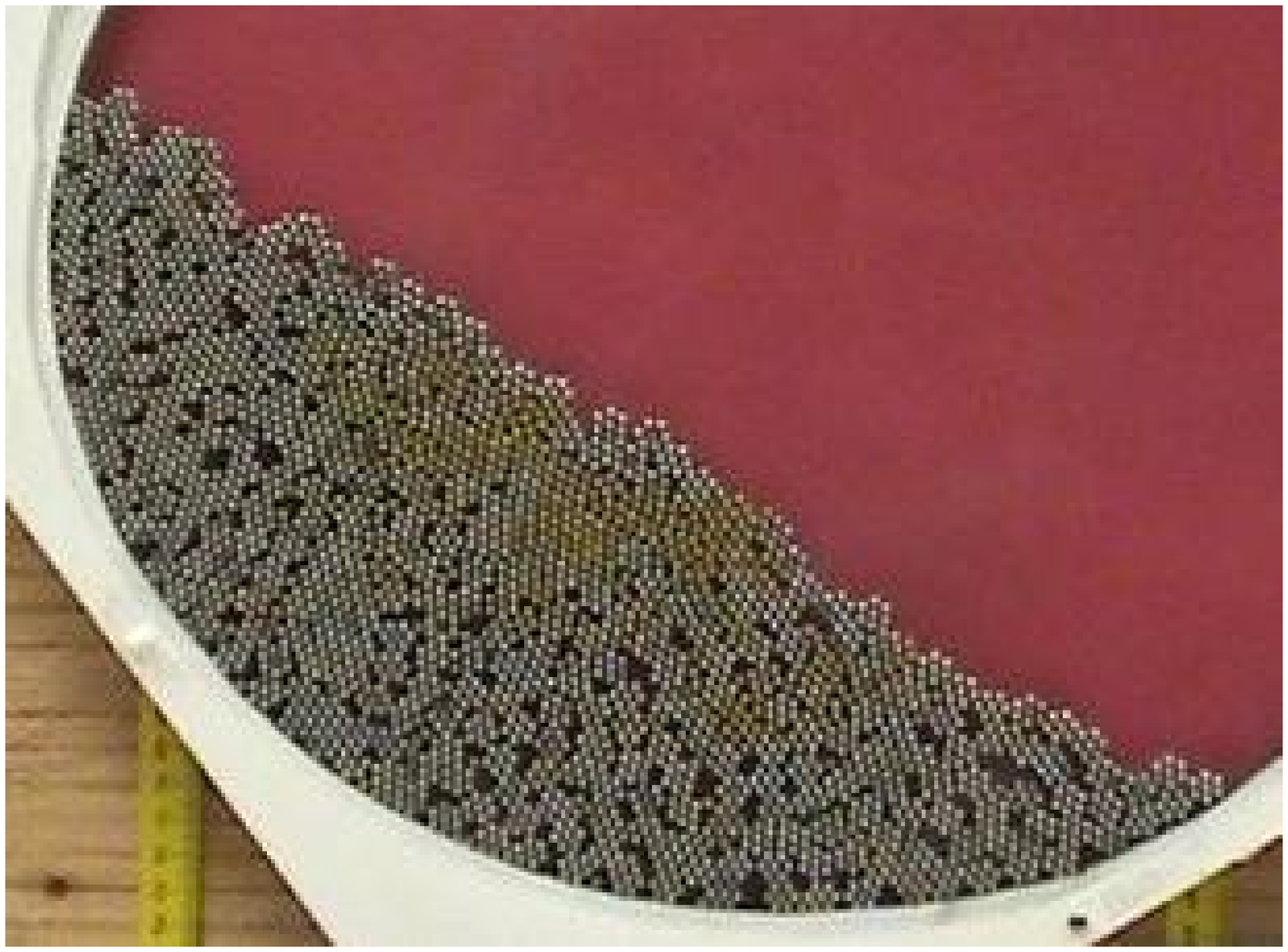}}\vspace{.06in}\\
\scalebox{.35}{\includegraphics{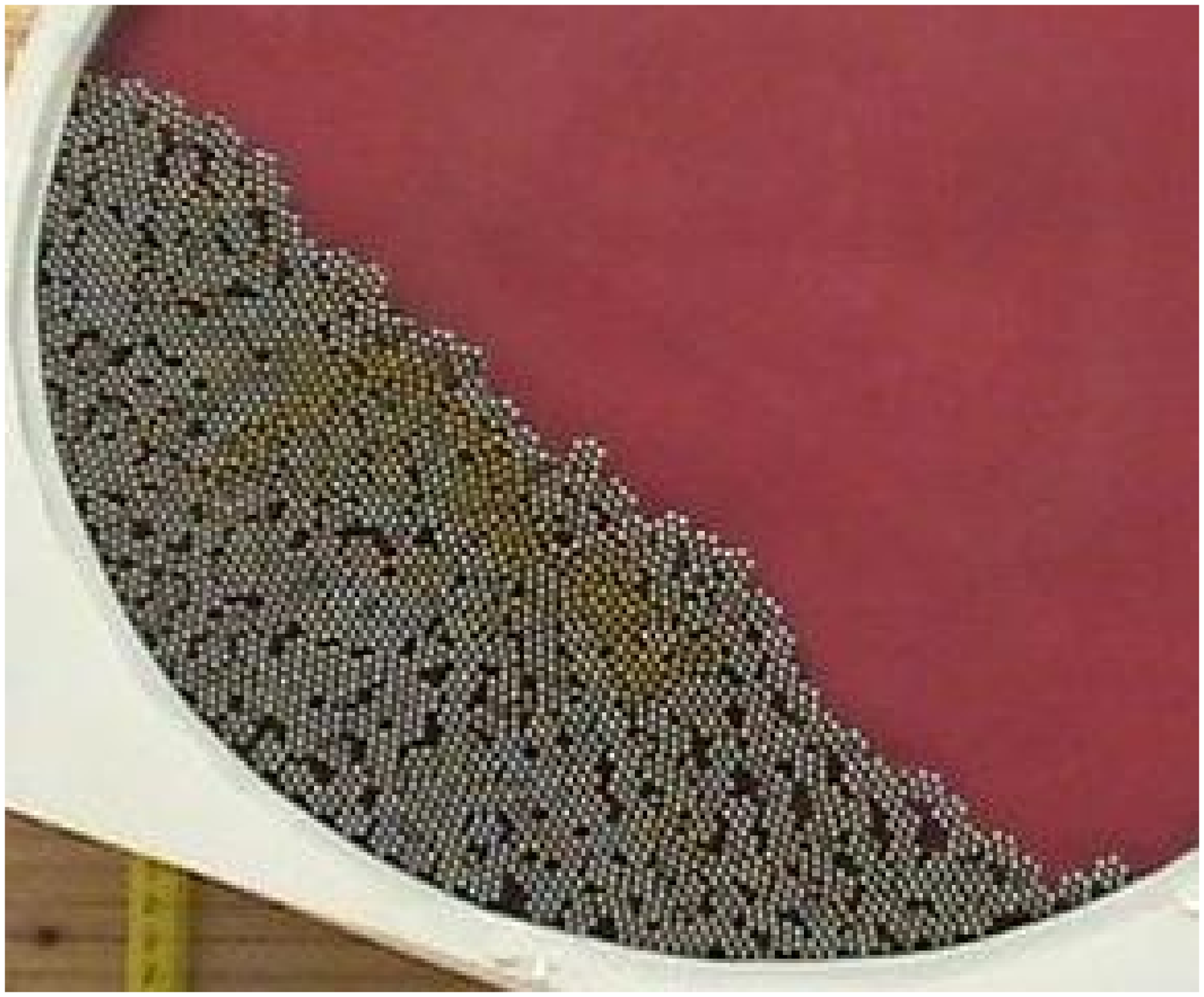}}\vspace{.06in}\\
\scalebox{.35}{\includegraphics{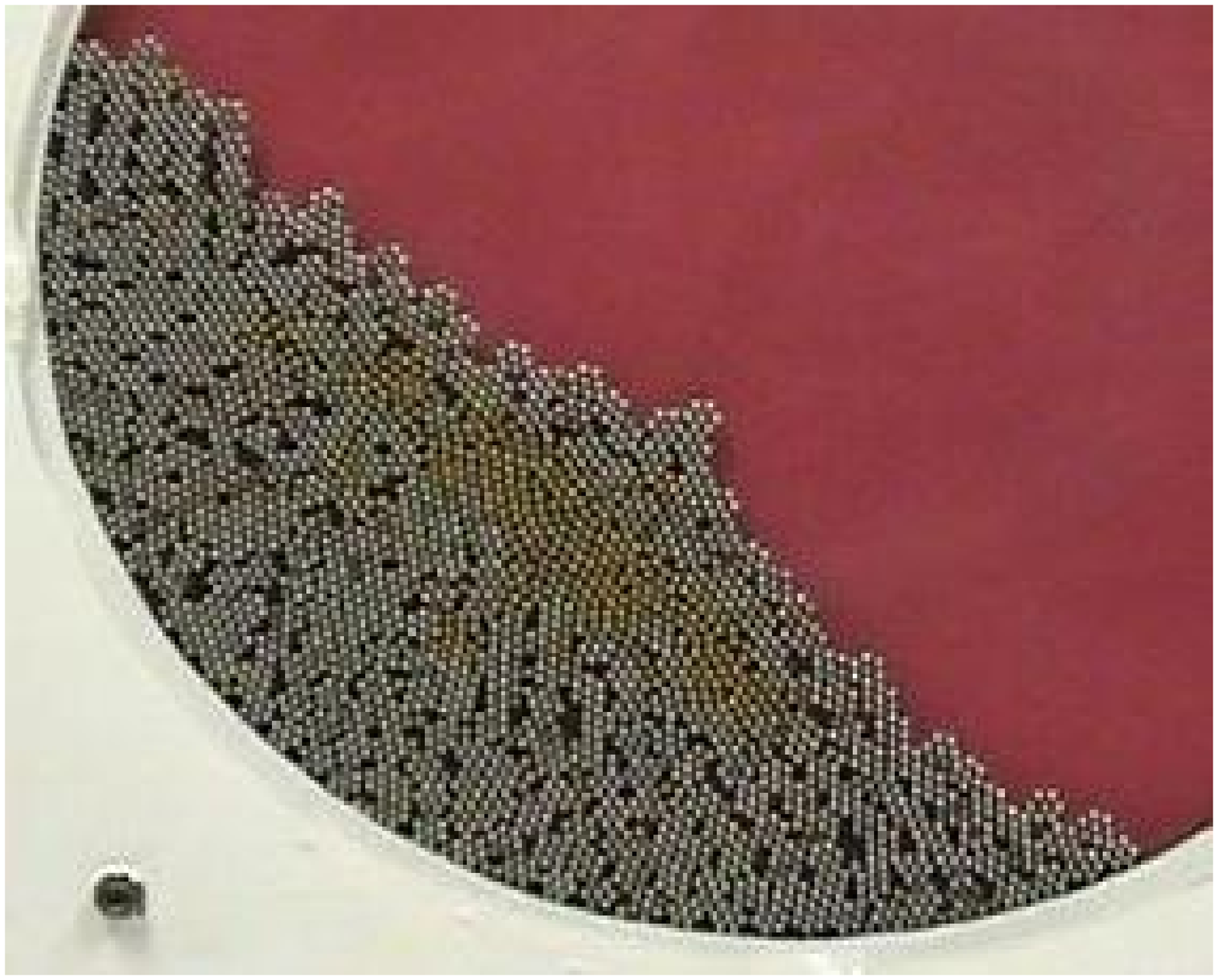}}
\caption{Images of Figure \ref{f:nearfar} with original colors.}
\label{f:nfcolor}
\end{figure}

Figures \ref{f:segcolor} and \ref{f:nfcolor} include the images of Figures
\ref{f:segregate} and \ref{f:nearfar} with their original colors.

\end{document}